\pdfoutput=1
\documentclass[aps,prl,reprint,superscriptaddress,nofootinbib,longbibliography]{revtex4-2}
\usepackage[T1]{fontenc}
\usepackage[utf8]{inputenc}
\usepackage{lmodern}
\usepackage{amsmath,amssymb,bm,booktabs}
\usepackage[colorlinks=true,citecolor=blue,linkcolor=blue,urlcolor=blue]{hyperref}
\hypersetup{pdftitle={Nature's simplest perfect strategy},pdfauthor={Emmanuel Zambrini Cruzeiro and Adán Cabello},pdfsubject={Bipartite pseudotelepathy: a fully analytical support theorem}}
\newcommand{\tr}{\operatorname{tr}}
\newcommand{\supp}{\operatorname{supp}}
\newcommand{\id}{I}
\newcommand{\norm}[1]{\left\lVert#1\right\rVert}

\newcommand{\Qc}{\mathcal Q_{\mathrm{qc}}}
\newcommand{\Qcl}{\overline{\mathcal Q}}
\begin{document}

\title{Nature's simplest perfect strategy}
\author{Emmanuel Zambrini Cruzeiro}
\affiliation{Quantum Information and Quantum Optics Laboratory, Instituto Superior T\'ecnico, Lisboa, Portugal}
\affiliation{Quantum Physics of Information Group, Instituto de Telecomunica\c{c}\~oes, Lisboa, Portugal}
\author{Ad\'an Cabello}
\affiliation{Departamento de F\'isica Aplicada II, Universidad de Sevilla, E-41012 Sevilla, Spain}
\affiliation{Instituto Carlos I de F\'isica Te\'orica y Computacional, Universidad de Sevilla, E-41012 Sevilla, Spain}

\begin{abstract}
Recently, S. Kumar \textit{et al.} [\textit{Science} \textbf{394}, 870 (2026)] reported the largest experimental quantum winning probability for a nonlocal game to date. They achieved this by preparing the magic-square correlation [\textit{Phys. Rev. Lett.} \textbf{87}, 010403 (2001)], which was conjectured to be the bipartite nonlocal correlation with the fewest inputs and outputs that allows for a perfect winning probability [\textit{Phys. Rev. Lett.} \textbf{134}, 010201 (2025)]. Here, we prove this conjecture by establishing a stronger result: no bipartite quantum correlation in which one party has three measurements, each with at most three outcomes, can achieve a perfect winning probability. This result, which holds for arbitrary states, measurements, and dimensions, establishes the fundamental role of the magic-square correlation in quantum theory and explains its ubiquity in quantum computation.
\end{abstract}

\maketitle

\textit{Introduction.---}Bell nonlocality shows that spacelike-separated observers can produce correlations that no local classical model can explain~\cite{Bell1964,CHSH1969}. 
Equivalently, a correlation violates an $N$-partite Bell inequality if and only if there exists a nonlocal game in which a team of $N$~mutually noncommunicating players equipped with such a correlation can win that game with higher probability than any local classical strategy \cite{Brunner2014}.

When this higher winning probability is one, the game admits a perfect quantum strategy, a situation known as quantum pseudo-telepathy~\cite{Brassard2005}. Such strategies are ubiquitous in recent theoretical and experimental results in quantum computation~\cite{Bravyi2018,Ji2021,Kumar2026}. This case is also especially interesting in physics since a quantum perfect strategy occurs if and only if \cite{Liu2024}: (i) the quantum correlation $p$ is fully nonlocal, meaning that the largest local fraction \(0 \le l_f \le 1\)~\cite{EPR1992} that can be assigned among all possible decompositions $p = l_f p_{L} + (1-l_f) p_{NS}$, where \(p_{L}\) is a local correlation and \(p_{NS}\) is a nonlocal nonsignaling correlation, is $l_f=0$; (ii) $p$ is not only outside the local polytope but actually in a face of the nonsignaling polytope that does not contain local points; and (iii) $p$ is in direct conflict with the Einstein--Podolsky--Rosen (EPR) criterion of elements of reality~\cite{EPR1935}: no local model can explain the predictions with certainty in $p$; equivalently, no local model can explain the zero components in $p$. 

Identifying the simplest quantum correlation with these equivalent properties is therefore a fundamental question in physics. In what follows, we refer to these properties in terms of perfect strategies.

It is known that, for $N \ge 3$ players, perfect strategies exist with two measurement settings of two outcomes~\cite{Mermin1990}. However, for $N=2$ players, no perfect quantum strategy can exist if at least one of the players has only two inputs~\cite{Gisin2007}. In fact, the first examples of bipartite perfect quantum strategies required hundreds of inputs~\cite{Stairs1983,HR1983}. The question then becomes: which bipartite perfect strategy uses the fewest inputs, and among those, which uses the fewest outputs? Building on the connection between perfect quantum strategies and Kochen--Specker sets~\cite{KS1967}, Ref.~\cite{Cabello2025} conjectured (Conjecture~1) that the simplest bipartite perfect quantum strategy is the magic-square correlation introduced in Refs.~\cite{Cabello2001,Cabello2001b} and later popularized in connection with the magic-square game~\cite{Aravind2002,Aravind2004} and perfect quantum pseudotelepathy~\cite{Brassard2005}.

The magic-square correlation requires three inputs and four outputs. How to produce it, and how it relates to the proof of the Kochen--Specker theorem for an indivisible system in Refs.~\cite{Peres1990,Mermin1990b}, is detailed in Appendix~\ref{app:magic}.

The magic-square correlation has been produced experimentally with pairs of photons hyperentangled in polarization and path~\cite{Cinelli2005,Yang2005,Barbieri2005}, following the proposal in~\cite{Chen2003}, in connection with the direct conflict with EPR---also referred to as an ``all-versus-nothing'' proof---and in connection with full nonlocality~\cite{Aolita2012}. It has also been generated with pairs of photons hyperentangled in polarization and orbital angular momentum in connection with pseudotelepathy~\cite{Xu2022}. Recently, it has also been prepared with superconducting circuits in connection with contextuality-based quantum computational advantage~\cite{Kumar2026}, achieving the largest experimental quantum winning probability for a nonlocal game to date, $98.30\pm0.01\%$, very close to the theoretical ideal of $100\%$.

The very existence of the magic-square correlation with three inputs per party, together with the proof~\cite{Gisin2007} that two inputs are not enough, implies that the remaining question is whether fewer outputs for some of the inputs suffice to produce a perfect strategy when both players have three inputs.

This question concerns the supports of quantum correlations, rather than the largest violation of a particular Bell inequality. Ref.~\cite{Liu2024} excluded perfect strategies in the Bell scenarios \((3,3,3,2)\) and \((3,3,2,4)\), where \((X,Y,A,B)\) indicates that Alice (Bob) has \(X\) (\(Y\)) inputs, each with \(A\) (\(B\)) outputs. Ref.~\cite{Cabello2025} excluded perfect strategies in \((3,3,3,3)\) for maximal observables, but left the unrestricted measurement problem open. Neither a restriction to rank-one projectors nor a fixed local dimension covers a general strategy: a three-outcome measurement may have effects of arbitrarily large rank, and a nonmaximally entangled state need not prepare orthogonal conditional states.

Here we prove a stronger result: if Alice has three inputs and at most three outputs, then no choice of Bob's finite input and output alphabets permits a perfect strategy. The proof uses the conditional states prepared on Bob's side, retaining their positivity and common average state without imposing orthogonality. A universal relation between their supports replaces the classification of individual forbidden-event patterns. The proof combines an inequality for a ternary ensemble with the equality conditions of an operator variance identity. The resulting theorem establishes the minimum output alphabets at the minimum input size and identifies the magic-square scenario as optimal.

\textit{Scenario and main result.---}A finite-dimensional quantum correlation has the form
\begin{equation}
p(a,b|x,y)=\tr\!\left[\rho_{AB}(A_a^x\otimes B_b^y)\right],
\label{eq:quantum}
\end{equation}
where $\rho_{AB}$ is a density operator and each measurement is a positive-operator-valued measure (POVM). A deterministic support assignment consists of functions $\alpha$ and $\beta$ such that
\begin{equation}
p\bigl(\alpha(x),\beta(y)\bigm|x,y\bigr)>0
\quad\text{for every }x,y.
\label{eq:assignment}
\end{equation}
If \(p\) wins a game perfectly, then every event with positive probability on a promised input pair is a winning event. Equation~\eqref{eq:assignment} then yields a perfect deterministic classical strategy. Conversely, if no such assignment exists, then \(p\) wins the game whose winning events are exactly its support, while no deterministic classical strategy does. This is the support formulation of correlations that allow for a perfect strategy~\cite{Liu2024}.

\textbf{Theorem 1.} \textit{Every finite-dimensional quantum correlation in the scenario $(3,m,3,n)$, for arbitrary finite $m,n$, admits a deterministic support assignment. The state may be mixed or pure, and all measurements may be arbitrary POVMs.}

The theorem includes $(3,3,3,3)$ and, by exchanging the players, every scenario in which either player has at most three inputs and at most three outputs. There is no restriction on Schmidt coefficients, local dimension, or the ranks of measurement effects. In particular, the statement applies to nonmaximally entangled states without converting them to maximally entangled ones.

\textit{Proof of Theorem 1.---}For each Alice input $x$, define Bob's subnormalized conditional states
\begin{equation}
\sigma_a^x=\tr_A\!\left[
(\sqrt{A_a^x}\otimes\id)\rho_{AB}
(\sqrt{A_a^x}\otimes\id)\right].
\label{eq:assemblage}
\end{equation}
They satisfy
\begin{equation}
\sigma_a^x\succeq0,\qquad
\sum_{a=0}^{2}\sigma_a^x=\rho_B,\qquad
p(a,b|x,y)=\tr(\sigma_a^xB_b^y).
\label{eq:ensembles}
\end{equation}
Alice's three measurements therefore give three ternary decompositions of the same density operator. Let $Q_a^x=\supp\sigma_a^x$, with $\supp0=0$. These are projectors onto the supports of Bob's conditional states and need not be mutually orthogonal or sum to the identity.

We use the following support lemma, the proof of which is given below. We first show that it yields the deterministic support assignment required by Theorem~1.

\textbf{Lemma 1 (Common-density support lemma).} \textit{For any three ternary decompositions in Eq.~\eqref{eq:ensembles}, there are outcomes $a_0,a_1,a_2$ such that}
\begin{equation}
\norm{Q_{a_0}^0+Q_{a_1}^1+Q_{a_2}^2}>2.
\label{eq:support-lemma}
\end{equation}

We now derive Theorem~1 from Lemma~1 using the connection between perfect strategies and antidistinguishability established in Ref.~\cite{Renner2026}, expressed here in terms of support projectors. We include the argument for completeness.

To see why this suffices, fix the three outcomes from the lemma and consider any Bob input $y$. Suppose that no Bob outcome is compatible with all three. For each$~b$, select an index $x(b)$ for which $p(a_{x(b)},b|x(b),y)=0$. Positivity in Eq.~\eqref{eq:ensembles} implies $Q_{a_{x(b)}}^{x(b)}B_b^y=0$. Group Bob's effects according to this index:
\begin{equation}
F_x=\sum_{b:x(b)=x}B_b^y,\qquad \sum_xF_x=\id.
\end{equation}
Since $Q_{a_x}^x F_x=0$ and $0\preceq F_x\preceq\id$, we have $F_x\preceq\id-Q_{a_x}^x$. Summing gives
\begin{equation}
Q_{a_0}^0+Q_{a_1}^1+Q_{a_2}^2\preceq2\id,
\end{equation}
contradicting Lemma~1. Hence, each Bob input has an outcome $b_y$ compatible with the same three Alice outcomes. Notice that the choice on Alice's side is independent of Bob's measurements. Setting $\alpha(x)=a_x$ and $\beta(y)=b_y$, we obtain
\begin{equation}
p\bigl(\alpha(x),\beta(y)\bigm|x,y\bigr)>0
\quad\text{for every }x,y.
\end{equation}
Thus $\alpha$ and $\beta$ form a deterministic support assignment. This completes the proof of Theorem~1 assuming Lemma~1.

\textit{Proof of Lemma 1.---}The proof rests on an inequality for a single ternary ensemble. Let $\rho=\sum_i\tau_i$ be a density operator, let $E_i=\supp\tau_i$, and put $X=\sum_iE_i$. We work on the support of $\rho$ and define $\varphi(T)=\tr(\rho T)$. Introduce the auxiliary polynomial
\begin{equation}
h(t)=t^3-8t^2+31t.
\end{equation}
We choose these coefficients to express the averaging identity below in terms of $h$ and obtain a lower bound whose equality condition forces $X=I$. Specifically, we will show that
\begin{equation}
\varphi(h(X))\geq24,
\quad\text{with equality only if }X=I.
\label{eq:main-cubic}
\end{equation}
Write $D=X-I$ and $S_i=X-E_i$. Since $E_i\tau_i^{1/2}=\tau_i^{1/2}$, the support relation gives
\begin{equation}
D\tau_i^{1/2}=S_i\tau_i^{1/2}.
\end{equation}
Consequently,
\begin{align}
d:=\varphi(D)&=\sum_i\tr(\tau_iS_i)\geq0, \label{eq:11} \\
e:=\varphi(D^2)&=\sum_i\tr(\tau_iS_i^2)\leq2d. \label{eq:12}
\end{align}
Inequalities \eqref{eq:11} and \eqref{eq:12} follow from $0\preceq S_i\preceq2I$, because $S_i$ is the sum of the other two projections; hence $S_i^2\preceq2S_i$. Also $e\geq0$. The polynomial identity
\begin{equation}
h(X)-24I=18D-6D^2+DXD
\end{equation}
and $DXD\succeq0$ give
\begin{equation}
\varphi(h(X))-24=18d-6e+\varphi(DXD)\geq6d\geq0.
\end{equation}
This proves Eq.~\eqref{eq:main-cubic}. Equality forces $d=e=0$. Since $\rho$ is strictly positive on its support, $e=\tr(\rho D^2)=0$ implies $D=0$ there, and therefore, $X=I$.

Write the three support families as $P_a,Q_b,R_c$, and suppose that all $L_{abc}=2I-P_a-Q_b-R_c$ are positive. Set $Y=\sum_bQ_b$, $Z=\sum_cR_c$, and
\begin{equation}
K_{bc}=4I+3Q_b-Y+3R_c-Z.
\end{equation}
Write $U_b=3Q_b-Y$ and $V_c=3R_c-Z$. These centered operators have zero uniform average over their three labels. For an arbitrary operator $H$ on Bob's Hilbert space, define the positive linear map
\begin{equation}
\begin{aligned}
\mathcal V_Q(H)&=\frac13\sum_bU_bHU_b\\
&=\sum_{b<d}(Q_b-Q_d)H(Q_b-Q_d).
\end{aligned}
\end{equation}
This is a finite operator-variance identity; see, for example, Corollary~2.2 of Ref.~\cite{MoslehianZhang2008}.
If $H\succeq0$, each term $U_bHU_b$ is positive because $U_b$ is self-adjoint. Thus $\mathcal V_Q$ preserves positivity. Define $\mathcal V_R$ analogously using $V_c$; we call these the variance maps of the two families.
Let $\alpha_a(T)=\tr(\sigma_a^0T)$. Expanding the nonnegative average $\mathcal E=\tfrac19\sum_{a,b,c}\alpha_a(K_{bc}L_{abc}K_{bc})$ gives
\begin{align}
0\leq\mathcal E\leq16-\tfrac13\varphi\bigl[&h(Y)+h(Z)\nonumber\\
&+\mathcal V_Q(Z)+\mathcal V_R(Y)\bigr]\leq0.
\label{eq:main-analytic-squeeze}
\end{align}
The expansion and the discarded nonnegative terms are given explicitly in Appendix~\ref{app:analytic}. The last inequality follows from Eq.~\eqref{eq:main-cubic}. Equality therefore forces $Y=Z=I$. But then $\mathcal V_Q(Z)=\mathcal V_R(Y)=2I$, and Eq.~\eqref{eq:main-analytic-squeeze} gives $0\leq\mathcal E\leq-4/3$, a contradiction.

Thus the assumption that all transversal support sums are at most $2I$ forces two ensembles to have orthogonal supports, and those supports already violate the same assumption. This proves Lemma~1 using only positivity and polynomial identities. Appendix~\ref{app:analytic} also gives a positive-functional formulation, which removes the finite-dimensional trace from the argument.

\textit{The minimum scenario.---}Combining Theorem~1 with the two-input exclusion gives an immediate consequence.

\textbf{Corollary 1.} \textit{Among bipartite games with a perfect finite-dimensional quantum strategy and no perfect classical strategy, the lexicographically minimum resource pair is}
\begin{equation}
(m_A m_B,\,o_A o_B)=(9,16),
\label{eq:minimum}
\end{equation}
\textit{which is achieved by the magic-square game in the $(3,3,4,4)$ Bell scenario.}

Indeed, the two-input theorem forces $m_A,m_B\geq3$~\cite{Gisin2007}; a support-based proof is included in Appendix~\ref{app:finite}. At the minimum $m_A m_B=9$, both parties have three inputs. Theorem~1 and its party-exchanged version then force $o_A,o_B\geq4$. The one-sided form is essential: excluding only $(3,3,3,3)$ would leave asymmetric output alphabets, such as $(3,3,3,4)$, undecided.

\textit{Correlation models and consequences.---}The proof of Theorem~1 directly covers finite-dimensional tensor-product strategies. However, its algebraic ingredient is stronger: the analytical proof uses positive functionals and their support relations, without requiring a trace. This permits a separate extension to commuting-operator strategies in arbitrary Hilbert spaces (Appendix~\ref{app:commuting}). The same deterministic-support conclusion therefore holds for the commuting correlation set $\Qc$ and, consequently, for the closure $\Qcl$ of finite-dimensional quantum correlations.

This extension also distinguishes exact attainment from a supremum of one. For every fixed finite game in $(3,m,3,n)$ without a perfect classical strategy, compactness of $\Qc$ gives $\omega_{\mathrm{qc}}<1$, where $\omega_{\mathrm{qc}}$ is the maximum commuting-operator winning probability. Thus, no sequence of finite-dimensional strategies can approach unit success for that fixed game. This establishes the existence of a game-dependent gap; it does not assign that gap a numerical value.

The support assignment also exhibits a positive local fraction $l_f$. For a given correlation, let $p_{L}$ be the deterministic correlation in Eq.~\eqref{eq:assignment} and choose $0<l_f<\min_{x,y}p(\alpha(x),\beta(y)|x,y)$. Then,
\begin{equation}
p_{NS}=\frac{p-l_f p_{L}}{1-l_f},
\end{equation}
where $p_{NS}$ is a nonsignaling correlation. Hence, the excluded scenarios cannot realize perfect strategies. 

\textit{Conclusions.---}We have shown that three ternary measurements on either party are not enough for a bipartite perfect strategy. Equivalently, they are not enough for full nonlocality, for correlations lying on a face of the nonsignaling polytope with no local points, or for correlations that directly contradict EPR reasoning. The equivalence is not merely terminological: each formulation stresses a different physical perspective. Our result shows that none of these three properties can be realized when one
side is restricted to three ternary measurements.

This holds independently of the number of measurements on the other party, the Hilbert-space dimension, and whether the shared state is pure or mixed, with no restriction on its entanglement spectrum. Moreover, since the obstruction is intrinsic to the measurement structure of one party, the result applies not only to finite-dimensional tensor-product models but also to all commuting-operator strategies.

The obstruction comes from a constraint on three ensembles with the same
average state. We prove it using two ingredients that may be of independent
interest: a cubic inequality for the sum of the support projections of a
ternary ensemble, and an averaged operator identity relating the supports of
the three ensembles. Together, they imply that every perfect quantum strategy
in these scenarios admits a perfect deterministic classical counterpart,
excluding perfect quantum strategies with quantum advantage. The proof uses operator positivity, polynomial identities, and the Cauchy--Schwarz inequality for positive functionals. These ingredients allow for the extension to commuting-operator strategies in arbitrary
Hilbert spaces established in Appendix~\ref{app:commuting}.

Combined with the two-input theorem, this result establishes the minimum
input and output alphabets for finite-dimensional bipartite perfect quantum strategies when inputs are minimized first. Each party requires at least three inputs. When both parties have three inputs, each must have at least one measurement with four or more outcomes. The magic-square correlation attains this minimum in $(3,3,4,4)$. In this precise sense, it is the simplest bipartite perfect nonclassical strategy allowed by quantum theory. This minimality explains why the magic-square, and its relatives such as the
Greenberger--Horne--Zeilinger--Mermin correlations are ubiquitous in quantum computation: they are the simplest ways to produce perfect quantum strategies.

\textit{Acknowledgments.---}A.C.\ is supported by the AEI/MICIU Project ``Quantum Correlations for Quantum Information and Computation'' (QCQIC) (Project No.\ PID2025-174467NB-I00). E.Z.C.\ was funded by national funds through FCT—Fundação para a Ciência e a Tecnologia, I.P., and, when eligible, co-funded by European Union funds under project UID/50008/2025—Instituto de Telecomunicações, DOI: 10.54499/UID/50008/2025. 

OpenAI ChatGPT (GPT-6 Astra) was used to assist with the exploration and refinement of candidate mathematical arguments and manuscript organization and editing. The authors specified the problems, algorithms, assumptions, and conventions; verified the resulting scientific claims and computational outputs; revised all AI-assisted material; and take full responsibility for the content of the manuscript.

\clearpage
\onecolumngrid
\appendix
\setcounter{secnumdepth}{2}

\section{Magic-square correlation}
\label{app:magic}

Here, we show how to produce the magic-square correlation \(p(a_{0|x},a_{1|x},b_{0|y},b_{1|y}|x,y)\), where \(a_{0|x},a_{1|x},b_{0|y},b_{1|y}\in\{+1,-1\}\) and \(x,y\in\{0,1,2\}\). This correlation enables a perfect quantum strategy in the \((3,3,4,4)\) Bell scenario \cite{Cabello2001b}, as it satisfies the winning condition of the magic-square game:
\begin{equation}
\label{mag}
\renewcommand{\arraystretch}{2}
\boxed{
\begin{array}{|c|c|c|}
\hline
{a_{0|0}}={b_{0|0}} &
{a_{1|0}}={b_{0|1}} &
{a_{0|0} a_{1|0}}={b_{0|2}}\\
\hline
{a_{0|1}}={b_{1|0}} &
{a_{1|1}}={b_{1|1}} &
{a_{0|1} a_{1|1}}={b_{1|2}}\\
\hline
{a_{0|2}}={b_{0|0} b_{1|0}} &
{a_{1|2}}={b_{0|1} b_{1|1}} &
{a_{0|2} a_{1|2}}={-b_{0|2} b_{1|2}}\\
\hline
\end{array}
}
\end{equation}

To produce this correlation, Alice and Bob share a maximally-entangled ququart-ququart state
\begin{equation}
|\Phi_4\rangle=\frac12\sum_{j=0}^3|j,j\rangle.
\end{equation}
Alice's measurements are given by
\begin{equation}
P_{a_0,a_1|x}=\frac14(I+a_0M_{x0})(I+a_1M_{x1}),
\end{equation}
and Bob's measurements are given by
\begin{equation}
Q_{b_0,b_1|y}=\left[\frac14(I+b_0M_{0y})(I+b_1M_{1y})\right]^{\mathsf T},
\end{equation}
where
\begin{equation}
(M_{xy})_{x,y=0}^2=
\begin{pmatrix}
Z\otimes I&I\otimes Z&Z\otimes Z\\
I\otimes X&X\otimes I&X\otimes X\\
Z\otimes X&X\otimes Z&Y\otimes Y
\end{pmatrix}
\end{equation}
is the Peres--Mermin table \cite{Peres1990,Mermin1990b}, with \(X,Y,Z\) the Pauli matrices, and \(I\) the identity.
For each \(x\), the four operators \(P_{a_0,a_1|x}\) are rank-one orthogonal projectors that sum to the identity; likewise, for each \(y\), the four operators \(Q_{b_0,b_1|y}\) are rank-one orthogonal projectors that sum to the identity.

With this state and these measurements, one obtains
\begin{equation}
p(a_{0|x},a_{1|x},b_{0|y},b_{1|y}|x,y)
=\frac14\operatorname{tr}\!\left(P_{a_0,a_1|x}Q_{b_0,b_1|y}\right)
=\frac{1+W_{xy}}{16},
\end{equation}
where
\begin{equation}
(W_{xy})_{x,y=0}^2=
\begin{pmatrix}
a_{0|0}b_{0|0} & a_{1|0}b_{0|1} & a_{0|0}a_{1|0}b_{0|2}\\[1mm]
a_{0|1}b_{1|0} & a_{1|1}b_{1|1} & a_{0|1}a_{1|1}b_{1|2}\\[1mm]
a_{0|2}b_{0|0}b_{1|0} & a_{1|2}b_{0|1}b_{1|1} & -a_{0|2}a_{1|2}b_{0|2}b_{1|2}
\end{pmatrix}.
\end{equation}
Therefore, \(W_{xy}=+1\) exactly for the winning events and \(W_{xy}=-1\) exactly for the losing events. Hence, every losing event has probability zero. Each of the eight winning events at a given input pair has probability \(1/8\). The quantum winning probability is one.

The magic-square correlation maximally violates a tight Bell inequality in the \((3,3,4,4)\) scenario \cite{Cabello2001b}, which can be written as
\begin{equation}\label{adanineq}
\begin{aligned}
\mu := {} &
\langle a_{0|0}b_{0|0} \rangle + \langle a_{1|0}b_{0|1} \rangle + \langle a_{0|0}a_{1|0}b_{0|2} \rangle \\
& + \langle a_{0|1}b_{1|0} \rangle + \langle a_{1|1}b_{1|1} \rangle + \langle a_{0|1}a_{1|1}b_{1|2} \rangle \\
& + \langle a_{0|2}b_{0|0}b_{1|0} \rangle + \langle a_{1|2}b_{0|1}b_{1|1} \rangle -\langle a_{0|2}a_{1|2}b_{0|2}b_{1|2} \rangle \le 7,
\end{aligned}
\end{equation}
where, e.g., \(\langle a_{0|0}b_{0|0} \rangle\) is the mean value of the product of \(a_{0|0}\) and \(b_{0|0}\). For the magic-square correlation,
\begin{equation}
\mu=9,
\end{equation}
which saturates the non-signaling and algebraic upper bound \(\mu\le 9\).

The accompanying program \texttt{verify\_magic\_square.py} independently checks the operator identities and probabilities described above using only integer matrices and rational arithmetic. It also examines all $4^6=4096$ deterministic strategies and finds $144$ strategies attaining the optimal classical winning probability $8/9$. No numerical tolerance is used.

\section{Conditional supports and the two-input obstruction}
\label{app:finite}

Here, we collect the elementary operator results used in the main text. Throughout, the Hilbert spaces are finite-dimensional, the state may be mixed, and measurement effects need not be projections.

\subsection{Conditional states and zero probabilities}

Equation~\eqref{eq:assemblage} defines a positive operator because partial trace preserves positivity. For any operator $T$ on Bob's space,
\begin{align}
\tr\!\left[T\sum_a\sigma_a^x\right]
&=\sum_a\tr\!\left[\rho_{AB}(A_a^x\otimes T)\right]
=\tr(\rho_B T).
\end{align}
Thus $\sum_a\sigma_a^x=\rho_B$ for every $x$. Taking $T=B_b^y$ gives the last identity in Eq.~\eqref{eq:ensembles}. This construction neither purifies the state nor dilates a POVM, so no conclusion about ranks or projectivity is needed.

For positive operators $S$ and $B$, write $Q=\supp S$. The following equivalence is useful:
\begin{equation}
\tr(SB)=0\quad\Longleftrightarrow\quad BQ=QB=0.
\label{eq:zero-support}
\end{equation}
Indeed, $\tr(SB)=\norm{B^{1/2}S^{1/2}}_2^2$. If it vanishes, $B^{1/2}$ annihilates the range of $S^{1/2}$, which is the range of $Q$. Conversely, $BQ=0$ implies $BS=0$. Here $\|\cdot\|_2$ is the Hilbert--Schmidt norm.

For clarity, if $Q$ is a projection, $0\preceq F\preceq I$, and $QF=0$, then
\begin{equation}
F=(I-Q)F(I-Q)\preceq I-Q.
\label{eq:effect-complement}
\end{equation}
Applying this to the three grouped effects in the main text proves the deterministic-support implication of Lemma~1. Grouping only establishes an operator inequality; it does not replace the original quantum strategy by a purportedly equivalent coarse-grained one.

For a game with input distribution $\pi$ and predicate $W\in\{0,1\}$, perfection means
\begin{equation}
\sum_{x,y,a,b}\pi(x,y)[1-W(x,y,a,b)]p(a,b|x,y)=0.
\end{equation}
All summands are nonnegative. Every losing event on an input pair with $\pi(x,y)>0$ therefore has zero probability, and a deterministic assignment contained in the support wins on every such pair. Pairs with $\pi(x,y)=0$ impose no constraint. Conversely, if no support assignment exists, taking all input pairs with positive weight and accepting precisely the positive events constructs a game with a perfect quantum strategy and no perfect classical one. Classical randomness cannot change this conclusion because a convex mixture has winning probability one only if every strategy of positive weight wins perfectly.

\subsection{Two measurements with arbitrary outcome numbers}

We give a direct common-density proof of the two-input obstruction of Ref.~\cite{Gisin2007}. Let $\{\sigma_a\}$ and $\{\tau_b\}$ be the conditional states for Alice's two measurements. Since both families sum to $\rho_B$,
\begin{equation}
\sum_{a,b}\tr(\sigma_a\tau_b)=\tr(\rho_B^2)>0.
\end{equation}
Some pair therefore has $\tr(\sigma_a\tau_b)>0$. Its support projections $P=\supp\sigma_a$ and $Q=\supp\tau_b$ are not orthogonal, by Eq.~\eqref{eq:zero-support}.

Suppose that a Bob measurement had no outcome compatible with both selected Alice outcomes. Group its effects according to which selected conditional state they exclude, choosing either group if both probabilities vanish. This gives $F+G=I$, $PF=0$, and $QG=0$. Equation~\eqref{eq:effect-complement} then yields $P+Q\preceq I$. For a vector $v$ in the range of $P$,
\begin{equation}
\langle v,Qv\rangle\leq
\langle v,(I-P)v\rangle=0,
\end{equation}
so $Qv=0$ and $P,Q$ are orthogonal, a contradiction. Every Bob measurement thus has a compatible outcome. The same selected pair works for all Bob inputs, yielding a deterministic support assignment. Measurements with one input can be duplicated to reduce to this case. No bound on either party's outcome alphabet is used.

\subsection{Scope of the resource minimum}

The output sizes $o_A,o_B$ are the maximum alphabet sizes over each party's inputs, with zero effects allowed. The two-input obstruction implies $m_A,m_B\geq3$ in any finite-dimensional bipartite game with a perfect strategy. Minimizing the Cartesian-product size $m_A m_B$ consequently gives nine and forces $m_A=m_B=3$. Theorem~1 then implies $o_A,o_B\geq4$, and the strategy in Appendix~\ref{app:magic} attains equality.

This proves optimality of the resource vector under the inputs-first ordering, including games with input promises. It does not count only the promised input pairs, assert uniqueness of an attaining game, or require every individual measurement of every attaining strategy to have four nonzero effects. Optimizing dimension or allowing additional inputs while minimizing outputs are different questions.

\section{Analytical proof of the support lemma}
\label{app:analytic}

Here, we prove an algebraic version of Lemma~1 using positivity, the Cauchy--Schwarz inequality for positive functionals, and explicit polynomial identities. No numerical calculation or external coefficient data enter the proof.

Let $P_a,Q_b,R_c$, with $a,b,c\in\{0,1,2\}$, be projections in a unital $C^*$-algebra. Suppose that positive linear functionals $\alpha_a,\beta_b,\gamma_c$ have a common normalized sum,
\begin{equation}
\sum_a\alpha_a=\sum_b\beta_b=\sum_c\gamma_c=\varphi,
\qquad\varphi(I)=1,
\label{eq:analytic-common}
\end{equation}
and are supported on the corresponding projections. For example,
\begin{equation}
\beta_b(Q_bT)=\beta_b(TQ_b)=\beta_b(T)
\label{eq:analytic-support}
\end{equation}
for every $T$, and similarly for $\alpha_a,P_a$ and $\gamma_c,R_c$. We show that some $P_a+Q_b+R_c$ has norm greater than two. The density-operator statement follows by taking $\alpha_a(T)=\tr(\sigma_a^0T)$, $\beta_b(T)=\tr(\sigma_b^1T)$, and $\gamma_c(T)=\tr(\sigma_c^2T)$. Orthogonality within a family is not assumed.

\subsection{A ternary-ensemble inequality and its equality case}

Consider one family $E_0,E_1,E_2$ with supported functionals $\eta_i$ summing to $\varphi$. Write
\begin{equation}
X=\sum_iE_i,\qquad D=X-I,\qquad S_i=X-E_i=\sum_{j\ne i}E_j.
\end{equation}
The support relations imply
\begin{align}
d:=\varphi(D)&=\sum_i\eta_i(S_i)\geq0,\label{eq:analytic-d}\\
e:=\varphi(D^2)&=\sum_i\eta_i(S_i^2)\leq2d.
\label{eq:analytic-e}
\end{align}
To obtain the equality in Eq.~(C5), substitute $D=(E_i-I)+S_i$ and expand the square. The support relation implies that the terms $(E_i-I)^2$, $(E_i-I)S_i$, and $S_i(E_i-I)$ vanish when evaluated by $\eta_i$, leaving $\eta_i(D^2)=\eta_i(S_i^2)$. The inequality in Eq.~(C5) follows from $0\preceq S_i\preceq2I$, which implies $S_i^2\preceq2S_i$, and the positivity of each $\eta_i$.

Set
\begin{equation}
h(t)=t^3-8t^2+31t.
\label{eq:analytic-h}
\end{equation}
Since $X\succeq0$, the polynomial identity
\begin{equation}
h(X)-24I=18D-6D^2+DXD
\end{equation}
and Eqs.~\eqref{eq:analytic-d}--\eqref{eq:analytic-e} give
\begin{equation}
\varphi(h(X))-24
=18d-6e+\varphi(DXD)
\geq6d\geq0.
\label{eq:analytic-cubic-bound}
\end{equation}
An explicitly positive form of the same identity is
\begin{equation}
\varphi(h(X))-24
=6\sum_i\eta_i(S_i)
+6\sum_i\eta_i\bigl((E_j-E_k)^2\bigr)
+\varphi(DXD),
\label{eq:analytic-positive-identity}
\end{equation}
where $\{j,k\}=\{0,1,2\}\setminus\{i\}$ in the $i$th summand. Indeed, $2S_i-S_i^2=(E_j-E_k)^2$. Equations~\eqref{eq:analytic-cubic-bound} and \eqref{eq:analytic-positive-identity} display explicitly where the three-outcome assumption enters.

The equality case will be essential. If $\varphi(h(X))=24$, then $d=0$. Every nonnegative summand in
\begin{equation}
d=\sum_i\sum_{j\ne i}\eta_i(E_j)
\end{equation}
must vanish. For $j\ne i$, Cauchy--Schwarz gives $\eta_i(E_jT)=\eta_i(TE_j)=0$, since $E_j^2=E_j$ and $\eta_i(E_j)=0$. Together with the support relation for $j=i$, this yields
\begin{equation}
\eta_i(E_jT)=\eta_i(TE_j)=\delta_{ij}\eta_i(T).
\label{eq:analytic-rigidity}
\end{equation}
Consequently, for every $T$,
\begin{equation}
\varphi\!\left(\sum_jE_jTE_j\right)=\varphi(T),
\qquad\varphi(XTX)=\varphi(T),
\qquad\varphi(X)=1.
\label{eq:analytic-equality-consequences}
\end{equation}
No faithful-state assumption is needed. In the finite-dimensional case with $\rho$ faithful on the working space, Eq.~\eqref{eq:analytic-e} also gives $\varphi((X-I)^2)=0$, and hence $X=I$.

\subsection{Averaging the forbidden transversal sums}

Assume, for contradiction, that every transversal sum has norm at most two. The 27 operators
\begin{equation}
L_{abc}=2I-P_a-Q_b-R_c
\label{eq:analytic-L}
\end{equation}
are then positive. Put $Y=\sum_bQ_b$ and $Z=\sum_cR_c$. Define
\begin{align}
\mathcal V_Q(H)
&=3\sum_bQ_bHQ_b-YHY\nonumber\\
&=\sum_{b<d}(Q_b-Q_d)H(Q_b-Q_d),
\label{eq:analytic-variance}
\end{align}
and define $\mathcal V_R$ analogously. These maps preserve positivity, as the last expression makes explicit.

For each $b,c$, set
\begin{equation}
K_{bc}=4I+3Q_b-Y+3R_c-Z.
\label{eq:analytic-K}
\end{equation}
Uniform averaging over the nine pairs $(b,c)$ gives the operator identity
\begin{align}
\frac19\sum_{b,c}K_{bc}L_{abc}K_{bc}
={}&32I-16P_a\nonumber\\
&-\frac13\left[h(Y)+h(Z)+\mathcal V_Q(Z)+\mathcal V_R(Y)\right]\nonumber\\
&-\mathcal V_Q(P_a)-\mathcal V_R(P_a).
\label{eq:analytic-average}
\end{align}
This is a polynomial identity, not an assumed inequality. It uses only $Q_b^2=Q_b$ and $R_c^2=R_c$; operators from different families need not commute.

To verify it by hand, write $U_b=3Q_b-Y$ and $V_c=3R_c-Z$, so that $K_{bc}=4I+U_b+V_c$. The required averages are
\begin{align}
\mathbb E_b U_b&=0,&
\mathbb E_b U_b^2&=3Y-Y^2,\nonumber\\
\mathbb E_b U_bHU_b&=\mathcal V_Q(H),&
\mathbb E_b U_bQ_bU_b&=\tfrac13Y(3I-Y)^2,\nonumber\\
\mathbb E_b(U_bQ_b+Q_bU_b)&=\tfrac23(3Y-Y^2),
\label{eq:analytic-centered}
\end{align}
with the corresponding formulas for $V_c$. Terms linear in a centered operator from the other family average to zero. Expanding $(4I+U_b+V_c)(2I-P_a-Q_b-R_c)(4I+U_b+V_c)$ using these formulas gives Eq.~\eqref{eq:analytic-average}. In particular, the $P_a$ contribution is $-16P_a-\mathcal V_Q(P_a)-\mathcal V_R(P_a)$.

\subsection{The contradiction}

Apply $\alpha_a$ to Eq.~\eqref{eq:analytic-average} and sum over $a$. Since $\alpha_a(P_a)=\alpha_a(I)$, the nonnegative quantity
\begin{equation}
\mathcal E=\frac19\sum_{a,b,c}\alpha_a(K_{bc}L_{abc}K_{bc})
\end{equation}
satisfies
\begin{align}
0\leq\mathcal E
={}&16-\frac13\varphi\!\left[h(Y)+h(Z)+\mathcal V_Q(Z)+\mathcal V_R(Y)\right]\nonumber\\
&-\sum_a\alpha_a\!\left[\mathcal V_Q(P_a)+\mathcal V_R(P_a)\right]
\leq16-\frac{24+24}{3}=0.
\label{eq:analytic-squeeze}
\end{align}
The last inequality uses positivity of the variance maps and Eq.~\eqref{eq:analytic-cubic-bound} for the $Q$ and $R$ ensembles. Equality must hold throughout; in particular,
\begin{equation}
\varphi(h(Y))=\varphi(h(Z))=24.
\end{equation}
Applying Eq.~\eqref{eq:analytic-equality-consequences} to each family now gives
\begin{align}
\varphi(\mathcal V_Q(Z))
&=3\varphi(Z)-\varphi(Z)=2,\nonumber\\
\varphi(\mathcal V_R(Y))&=2,
\end{align}
because $\varphi(Y)=\varphi(Z)=1$. Keeping these two contributions in Eq.~\eqref{eq:analytic-squeeze} yields
\begin{equation}
0\leq\mathcal E\leq16-\frac{24+24+2+2}{3}=-\frac43,
\end{equation}
a contradiction. Therefore some $L_{abc}$ is not positive, proving the support lemma.

The argument uses arbitrary positive functionals and therefore applies in the commuting-operator setting of Appendix~\ref{app:commuting}. In addition, the first family $\{P_a,\alpha_a\}$ was only summed using Eq.~\eqref{eq:analytic-common}; its number of outcomes was never used. Thus the same result holds if two of Alice's three measurements have at most three outcomes and the remaining measurement has an arbitrary finite number of outcomes.

\section{Commuting-operator strategies and closure}
\label{app:commuting}

We now extend Theorem~1 to commuting-operator strategies. Consider POVMs $\{A_a^x\}$ and $\{B_b^y\}$ on an arbitrary Hilbert space, with
\begin{equation}
[A_a^x,B_b^y]=0,\qquad
p(a,b|x,y)=\langle\psi,A_a^xB_b^y\psi\rangle,
\end{equation}
where $\psi$ is a unit vector. The vector formulation covers arbitrary algebraic states by their Gelfand--Naimark--Segal representation. Let $\mathcal M$ be the von Neumann algebra generated by Bob's effects. Then Alice's effects and their positive square roots belong to its commutant $\mathcal M'$.

For $v_a^x=\sqrt{A_a^x}\psi$, define positive normal functionals on $\mathcal M$ by
\begin{equation}
\varphi_a^x(T)=\langle v_a^x,Tv_a^x\rangle.
\end{equation}
Commutation and $\sum_aA_a^x=I$ give
\begin{equation}
\sum_a\varphi_a^x(T)=\langle\psi,T\psi\rangle=\varphi(T),
\qquad \varphi(I)=1,
\label{eq:commuting-common}
\end{equation}
independently of $x$. The support projections needed by the analytical lemma can be constructed inside $\mathcal M$. Define $Q_a^x$ as the orthogonal projection onto
\begin{equation}
\mathcal H_a^x=\overline{\{Cv_a^x:C\in\mathcal M'\}}.
\label{eq:cyclic-support}
\end{equation}
Here, the overline denotes closure in the Hilbert-space norm. Thus $\mathcal H_a^x$ includes all norm limits of vectors of the form $Cv_a^x$, and $Q_a^x$ is the orthogonal projection onto this closed subspace. The subspace is invariant under every element of the self-adjoint algebra $\mathcal M'$ and its adjoint, so it reduces $\mathcal M'$. Consequently, $Q_a^x\in(\mathcal M')'=\mathcal M$. Since $Q_a^xv_a^x=v_a^x$,
\begin{equation}
\varphi_a^x(Q_a^xT)=\varphi_a^x(TQ_a^x)=\varphi_a^x(T).
\label{eq:commuting-endpoints}
\end{equation}
Equations~\eqref{eq:commuting-common} and \eqref{eq:commuting-endpoints} are precisely the hypotheses of the positive-functional lemma in Appendix~\ref{app:analytic}. Some selected triple of projections therefore has sum with norm greater than two.

It remains to connect these projections to zero probabilities. If $B\succeq0$ belongs to $\mathcal M$ and $\varphi_a^x(B)=0$, then $\sqrt Bv_a^x=0$. Since $\sqrt B$ commutes with every $C\in\mathcal M'$, we have
\begin{equation*}
\sqrt B\,Cv_a^x=C\sqrt B\,v_a^x=0.
\end{equation*}
Thus $\sqrt B$ maps every vector of the form $Cv_a^x$ to zero. These vectors are dense in $\mathcal H_a^x$ in the Hilbert-space norm. Since $\sqrt B$ is bounded, continuity implies that it maps every vector in this closed subspace to zero. Thus,
\begin{equation}
\varphi_a^x(B)=0\quad\Longrightarrow\quad BQ_a^x=Q_a^xB=0.
\end{equation}
The grouping argument in the main text now applies in $\mathcal M$ and provides a deterministic support assignment. This proves the commuting-operator extension without using a finite-dimensional trace or assuming a tensor-product representation.

For a fixed finite Bell scenario, $\Qc$ is compact~\cite[Proposition~3.4]{Fritz2012}. To see this, consider the universal unital $C^*$-algebra generated by the two families of positive contractions, their POVM normalization relations, and cross-party commutation. Its state space is weak-* compact, and the evaluation map
\begin{equation*}
\omega\longmapsto\bigl(\omega(A_a^xB_b^y)\bigr)_{a,b,x,y},
\end{equation*}
which sends each algebraic state to its Bell probability table, is continuous. Conversely, the Gelfand--Naimark--Segal representation of every state yields a commuting strategy. Hence the image of this evaluation map is exactly~$\Qc$. Every finite-dimensional tensor-product correlation belongs to it, and therefore,
\begin{equation}
\Qcl\subseteq\Qc.
\end{equation}
Consequently, in $(3,m,3,n)$ every correlation in either set admits a deterministic support assignment.

Finally, fix a finite game in this scenario whose classical value is less than one. Its winning probability is a continuous linear functional on $\Qc$, so it attains a maximum. A maximum equal to one would supply a perfect commuting-operator strategy. By the commuting-operator extension of Theorem~1 proved above, this strategy admits a deterministic support assignment and therefore yields a perfect deterministic classical strategy. Therefore,
\begin{equation}
\omega_{\mathrm{qc}}<1.
\end{equation}
In particular, the finite-dimensional quantum supremum is also strictly less than one. This conclusion is made separately from the finite-dimensional theorem, which by itself excludes exact attainment only. The argument establishes a strict gap for each fixed finite game, including its chosen input distribution; it does not give a numerical value of that gap.

\end{document}